# Investigating the importance of edge-structure in the loss of H/H$_2$ of PAH cations: the case of dibenzopyrene isomers

Sarah Rodriguez Castillo[1,2], Aude Simon[2], Christine Joblin[1*]

1 Institut de Recherche en Astrophysique et Planétologie IRAP, Université de Toulouse (UPS), CNRS, CNES, 9 Av. du Colonel Roche, 31028 Toulouse Cedex 4, France
3 Laboratoire de Chimie et Physique Quantiques LCPQ/IRSAMC, Université de Toulouse (UPS) and CNRS, 118 Route de Narbonne, F-31062 Toulouse, France
*: christine.joblin@irap.omp.eu.

We present a detailed study of the main dehydrogenation processes of two dibenzopyrene cation (C$_{24}$H$_{14}^+$) isomers, namely dibenzo(a,e)pyrene (AE$^+$) and dibenzo(a,l)pyrene (AL$^+$). First, action spectroscopy under VUV photons was performed using synchrotron radiation in the 8-20 eV range. We observed lower dissociation thresholds for the non-planar molecule (AL$^+$) than for the planar one (AE$^+$) for the main dissociation pathways: H and 2H/H$_2$ loss. In order to rationalize the experimental results, dissociation paths were investigated by means of density functional theory calculations. In the case of H loss, which is the dominant channel at the lowest energies, the observed difference between the two isomers can be explained by the presence in AL$^+$ of two C-H bonds with considerably lower adiabatic dissociation energies. In both isomers the 2H/H$_2$ loss channels are observed only at about 1 eV higher than H loss. We suggest that this is due to the propensity of bay H atoms to easily form H$_2$. In addition, in the case of AL$^+$, we cannot exclude a competition between 2H and H$_2$ channels. In particular, the formation of a stable dissociation product with a five-membered ring could account for the low energy sequential loss of 2 hydrogens. This work shows the potential role of non-compact PAHs containing bay regions in the production of H$_2$ in space.

## 1. INTRODUCTION

The role of Polycyclic Aromatic Hydrocarbons (PAHs) as catalysts for $H_2$ formation has focused a lot of attention in particular because of its impact in astrochemistry [1–3]. Different mechanisms have been proposed, all involving the adsorption of at least one H atom. The Eley-Rideal mechanism considers first that a gas-phase hydrogen is attached to a PAH, preferably on an edge site, and is then abstracted by a passing-by H, forming $H_2$ [4–6]. Other studies [7,8] consider the hydrogenation of two or more carbon atom sites. These aliphatic carbons are then responsible for forming and ejecting $H_2$. For regular PAHs photo-fragmentation is found in experiments to be dominated by H loss [9,10], whereas some studies report the potential importance of $H_2$ loss but at higher energies than the threshold observed for H loss [11]. This is also comforted by static density functional theory (DFT) calculations [12].

The description of the hydrogenation state of PAHs in recent astrophysical models [13] has been based on the study of a single compact PAH, the coronene cation ($C_{24}H_{12}^+$). In a similar manner to coronene, pyrene ($C_{16}H_{10}^+$) has been shown to fragment mainly by H atoms [14]. However, in order to improve the astrophysical models, the family of PAH species to be considered should include more diversity, in particular in the geometry. In this work we study the fragmentation of the dibenzo-pyrene species, containing the same number of carbons as coronene but having a less compact structure. Structures containing bay areas have been shown to exhibit characteristic infrared signatures. One is located at shorter wavelength than the usual C-H stretch mode due to the steric interaction between the hydrogen atoms that are across bay regions [15]. Another one is located at 12.7 µm and is a good candidate to account for the band observed at this position in astrophysical spectra [16]. To further explore the influence of the structure on fragmentation we performed action spectroscopy experiments under VUV photons on two isomer cations: dibenzo(a,e)pyrene and dibenzo(a,l)pyrene (hereon referred to as $AE^+$ and $AL^+$ respectively). We analyze here in detail these experimental results by performing DFT calculations to evaluate the structures and paths involved in the dissociation of both isomers. We discuss in particular the competition between 2H and $H_2$ channels.

## 2. METHODOLOGY

### 2.1. Experimental methods

The two samples of AE and AL ($C_{24}H_{14}$) were obtained from the PAH Research Institute in Greifenberg (Dr. Werner Schmidt). These two molecules contrast in structure, differing in the location of one of the benzene rings attached to the pyrene skeleton (see Figure 1). AE presents a planar geometry, with a more open structure while AL is more compact, with a non-planar configuration due to the steric repulsion of two hydrogens ($H_f$ and $H_g$) located in adjacent benzene rings.

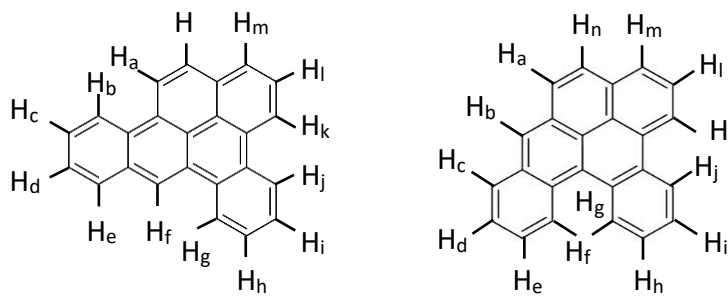

**Figure 1. Schematic geometries of dibenzo(a,e)pyrene cation ($AE^+$) -left- and dibenzo(a,l)pyrene cation ($AL^+$) -right- with labelling used in the manuscript.**

These molecules were part of a set of eight PAH samples that we subjected to synchrotron VUV irradiation at the SOLEIL facility in Saint-Aubin (France). The photo-processing of these samples was determined using action spectroscopy with the linear trap quadrupole (LTQ) setup available in the DESIRS beamline [17]. A complete description of the experimental campaign can be found in a previous paper [18], where we report the overall competition between the second ionization and fragmentation processes for all the studied species. Briefly, the main characteristics of DESIRS are a high photon flux, maintained in the range of $10^{12}$ to $10^{13}$ photons·s$^{-1}$, and a highly focused beam, in which the high harmonics are filtered out by rare gases depending on the energy range. The PAH neutral samples are charged in an atmospheric pressure photoionization source, where a nebulized solution of toluene and PAH is exposed to a 10.6 eV discharge from a krypton discharge lamp providing the source of ionization. The PAH cations stored in the ion trap are thermalized at room temperature (293 K) by collisions with helium atoms before photoprocessing. Synchrotron photons reach the ion trap (pressure ~$10^{-3}$ mbar) during a certain irradiation time, controlled by a beam shutter and selected to obtain the best signal. Two sets of data were obtained for each PAH cation: one for the low incident photon energies (8.0 − 15.6 eV with steps of 0.3 eV) and one for the high photon energies (15 – 20.0 eV with steps of 0.5 eV). The corresponding irradiation times were optimized to 1.0 and 0.2 s respectively for the low and high energy ranges; the synchrotron light can be considered as continuous in our experimental conditions. Photo-products were recorded as a function of the incident photon energy. Mass-spectra were averaged around 150 (400) times for low (high) energies. Blank signals recording the background were also obtained for each photon energy in similar circumstances as the PAH signals, for both isomers. These were subtracted from the full signal in order to determine the intensity of the peaks in the mass spectrum over the 3 sigma level.

## 2.2. Theoretical methods

In order to rationalize experimental results, the most direct dissociation pathways for AE$^+$ and AL$^+$ were computed at the DFT level of theory. The elementary steps - reaction intermediates and transition states - were investigated at the B3LYP/6-31G(d,p) level of theory using the Gaussian09 Package [19]. Local geometry optimizations were complemented by full harmonic frequency calculations by diagonalizing the Hessian matrix at this level. Each stationary geometry was characterized as a minimum or a saddle point of first order by frequency calculations, which were also used to obtain the zero-point vibrational energies (ZPEs). In the following, all mentioned energy values include ZPEs unless explicitly stated. Transitions states (TS) were optimized using the interpolation methods implemented in the Gaussian09 suit of programs. The Berny algorithm was used for most elementary steps. We used in a few cases the Synchronous Transit-Guided Quasi-Newton (STQN) algorithm and in litigious cases, the initial TS search was complemented with the Intrinsic Reaction Coordinate (IRC) procedure to confirm the reaction path.

Two possible spin-states were considered for the fragment ions, triplet and singlet for (PAH-H)$^+$, doublet and quartet for (PAH-2H)$^+$. Regarding singlet-spin states, only closed-shell systems were computed -in the restricted formalism- unless explicitly mentioned. For doublet and higher spin-states, the calculations were achieved within the unrestricted formalism, and expectation value of the $\widehat{S}^2$ spin operator was checked in order to detect potential spin-contamination.

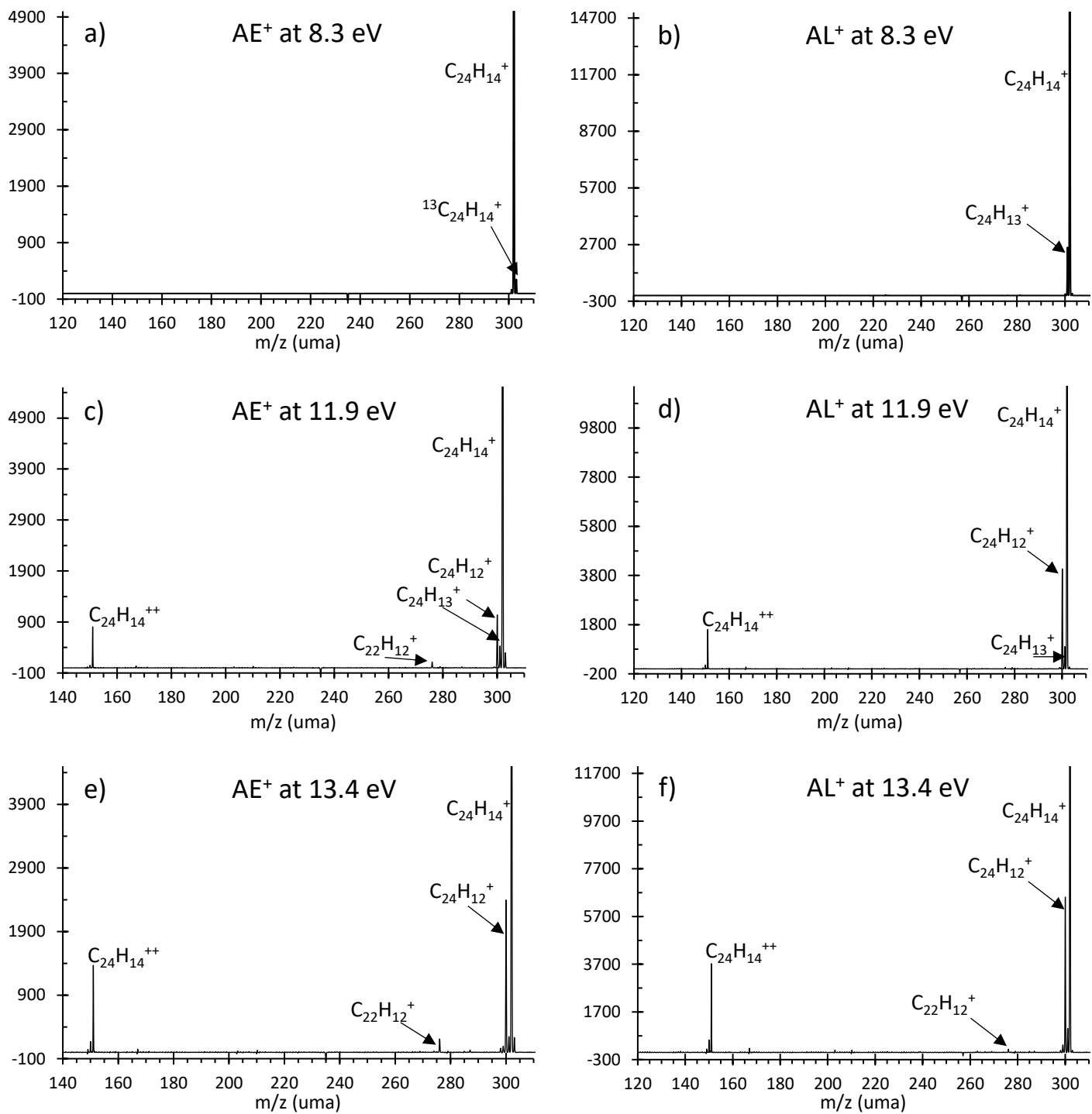

**Figure 2.** Mass spectra obtained with the LTQ mass spectrometer for AE$^+$ (left) and AL$^+$ (right) under different VUV irradiation energies (8.3, 11.9 and 13.4 eV). The parent is labeled by $C_{24}H_{14}^+$. $C_{24}H_{13}^+$ and $C_{24}H_{12}^+$ correspond to the loss of H and 2H/H$_2$ fragments, respectively. $C_{24}H_{14}^{++}$ corresponds to the doubly ionized parents. The vertical axis represents the intensities of the processed signal and it has been scaled in each case to 1/5 of the parent peak value.

## 3. RESULTS AND DISCUSSION
### 3.1. Experimental results

After subjecting the isomer cations to VUV photons of energies ranging from 8 to 20 eV, the dissociation products include the loss of up to four hydrogen atoms, together with carbon fragments, mainly acetylene. Three examples of processed spectra for each isomer can be seen in Figure 2, in increasing order of incident photon energies. For the lower energies (8.3 eV) no fragmentation is observed for $AE^+$ (Figure 2a), but a clear peak corresponding to H loss is detected for $AL^+$ (Figure 2b). Continuing up in photon energy, at 11.9 eV, the second ionization threshold has been surpassed and peaks corresponding to $C_{24}H_{14}^{++}$ are distinguishable, together with single and double hydrogen losses for both isomers and a small peak corresponding to $C_2H_2$ loss in the case of $AE^+$ (Figures 2c and 2d). The intensity of the H losses is visibly higher for $AL^+$ than for $AE^+$. The last example is at 13.4 eV, where both isomers present peaks with similar intensities, with the sole difference concerning the $C_2H_2$ loss, considerably more abundant in the case of $AE^+$ (Figures 2e and 2f).

The intensity of all the fragments normalized to the total number of ions is depicted in Figure 3 as a function of energy (left for $AE^+$, right for $AL^+$). The first channel to open is the loss of H: in both cases this path remains under 0.5%, and it decreases to almost 0% after 16 eV. At higher energies the dominant fragmentation process is the loss of two hydrogens (2H and/or $H_2$; the experimental technique cannot discriminate the two pathways), increasing almost linearly for both isomers up to an abundance of 2.4% at 13 eV, stabilizing up to around 17 eV, and decaying again to 0.2% at the highest energy, 20 eV. Other fragments involving carbon losses are gathered together under the $C_xH_y$ label (x= [1, 2], y= [2, 4]). In the following, we focus on low dissociation energies, at which these $C_xH_y$ channels are negligible.

Figure 4 zooms on the lowest photon energies, where essential differences can be observed in the -H and -2H/$H_2$ fragments, both on the dissociation thresholds and on the curve shape. In particular, Figure 4b emphasizes the difference in the 2H/$H_2$ channel below 12.8 eV. Quantifying dissociation energies at threshold is tricky, and in addition we have missed this threshold for the H loss of $AL^+$. We therefore decided to derive experimental onset energies for a fragment abundance of 0.1%, which corresponds to the value of the -H fragment abundance at the lowest photon energy recorded (8 eV). We derived the corresponding abundances in the case of $AE^+$ by considering that the number of ions that have dissociated is proportional to the number of ions that have absorbed a photon. Based on calculated values of the absorption cross-sections [18], we obtained a mean correction factor of 0.97 ± 0.03 for the [8 - 11 eV] range and for $AE^+$ relative to $AL^+$. This correction adds an extra 0.1 eV to the errors in the case of $AE^+$. In both isomers a typical error of 0.1 eV is due to extrapolation between two experimental points. For the loss of H we obtained onset energies of 8.0 ± 0.1 eV in the case of $AL^+$ and 10.2 ± 0.2 eV in the case of $AE^+$ (Figure 4a). This implies an energy difference of 2.2 ± 0.2 eV. Regarding 2H/$H_2$ loss, this led to values of 9.3 ± 0.1 eV and 10.8 ± 0.2 eV respectively (Figure 4b), giving an energy difference of 1.5 ± 0.2 eV.

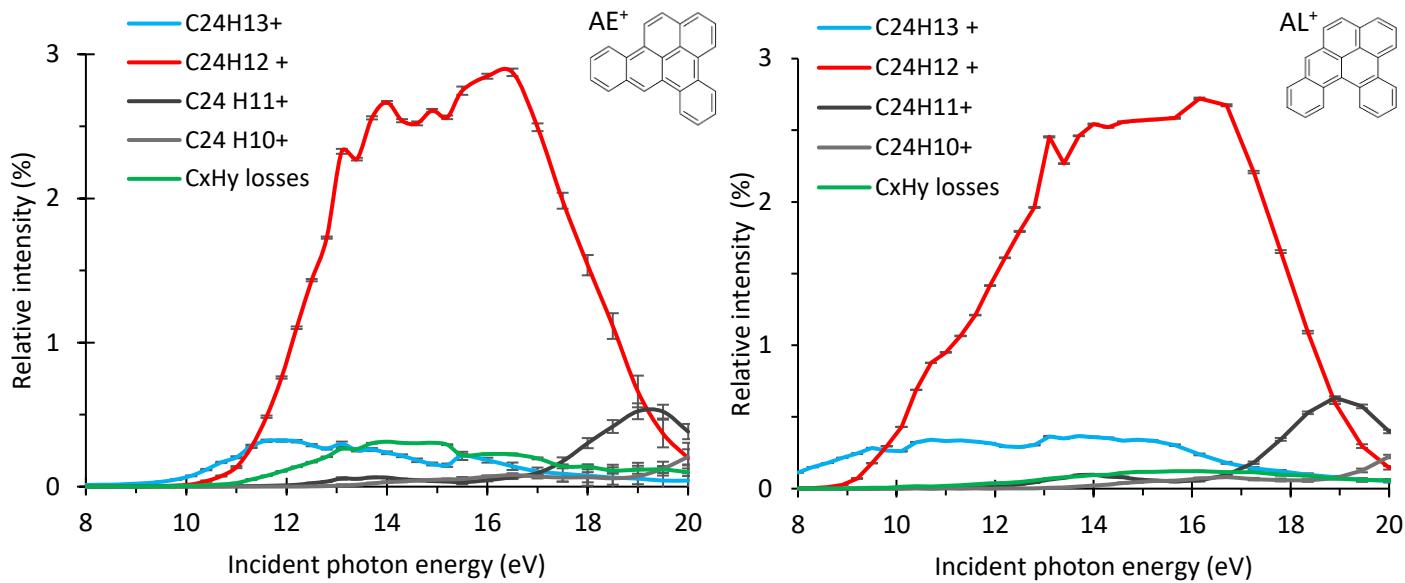

**Figure 3.** Abundances of the photofragments of AE$^+$ (left) and AL$^+$ (right) as a function of the VUV photon energy. The values are normalized to the total number of trapped ions and estimated error bars are also reported. Up to 4 hydrogen losses are observed: loss of H (C$_{24}$H$_{13}^+$) (blue), loss of 2 hydrogens (C$_{24}$H$_{12}^+$) (red), loss of three hydrogens (C$_{24}$H$_{11}^+$) (dark grey) and loss of 4 hydrogens (C$_{24}$H$_{10}^+$) (light grey). All fragments involving carbon loss are summed up in green.

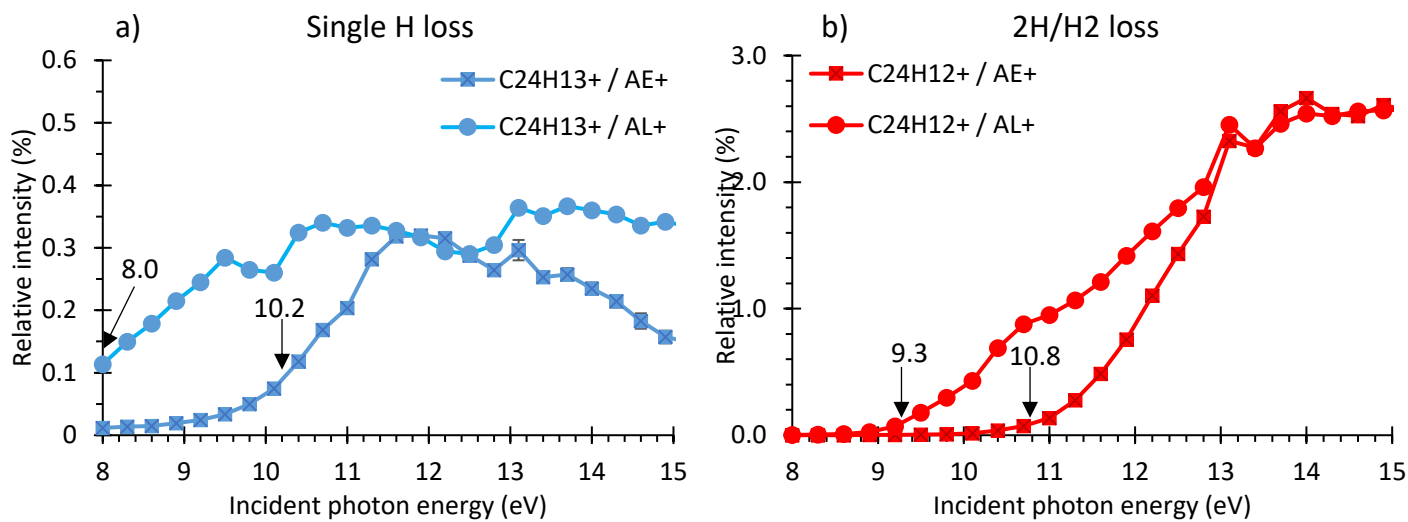

**Figure 4.** Left: zoom on the abundance of C$_{24}$H$_{13}^+$ formed by H loss as a function of the VUV photon energy. Right: similar for C$_{24}$H$_{12}^+$ resulting from 2H/H$_2$ loss. Circles correspond to AL$^+$ and squares to AE$^+$ parent cations. Error bars are included in the boxes.

### 3.2. Theoretical results

DFT calculations were carried out to investigate three possible different dehydrogenation paths: i) direct loss of H, ii) sequential loss of two H atoms and iii) loss of H$_2$ after the migration of a hydrogen to an adjacent position, the latter mechanism having been shown as the most energetically favorable path for H$_2$ loss [12]. Regarding the parent ions, the AL$^+$ ($^2$A electronic state) isomer was found 0.3 eV above AE$^+$ ($^2$A'), which could be expected because of steric hindrance, as mentioned in Section 2.1. In the following we will use the term "theoretical dissociation energy" for the difference between the enthalpies at 0K (electronic energy + ZPE) of the optimized separate products and reactants, and quoted hereafter ΔH(0K).

### 3.2.1. Direct loss of H

For the loss of a single hydrogen we computed the direct loss of H, since such rapid mechanism is likely to occur at the low energies at play. Molecular dynamic calculations [20] have confirmed that H loss is a fast and direct process. Nevertheless, other mechanisms involving isomerization with competing energetic barriers such as the ones suggested in [21] cannot be excluded, but a comprehensive study is out of reach with the approach used in the present paper.

None of the H atoms being equivalent, we considered the removal of H from all the possible sites in the initial geometry. Two possible spin-states (singlet and triplet) were investigated for the dehydrogenated products. In the following, we discuss the results for each isomer separately. Values of dissociation energies for all fragment isomers for molecular cations as well as the $\hat{S}^2$ eigenvalues can be found in Table 1 in the Annex.

In the case of $AE^+$ all the dehydrogenated products in the triplet spin-states could be easily optimized whereas closed-shell singlet spin-states could only be obtained for three sites (-$H_a$, -$H_f$, -$H_n$), the others leading to convergence problems despite the use of Fermi temperatures and quadratic convergence algorithm. In the cases in which convergence was achieved, the Singlet-Triplet gap was found in the [0.71 - 0.77 eV] range. This positive gap is in line with the results from [22,23]. The dehydrogenation dissociation energies lie in the [4.67 - 4.89 eV] range for these triplet-spin state products. These values are in good agreement with the results presented for the coronene cation ($C_{24}H_{12}^+$) by [12], whereas [11] obtained a value of 5.24 eV using a different basis set.

For such large conjugated systems where triplet spin-states are stable, open-shell singlet spin states are likely to compete. These could be obtained by multireference calculations that are prohibitive for our systems. An alternative to compute singlet diradicals with unrestricted DFT methods is to perform broken-symmetry (BS) calculations. We searched for this BS ms=0 solution for the $AE^+$-$H_f$ isomer. In addition, using a simplified procedure to estimate the spin-decontaminated singlet geometry [24,25] we obtained a singlet-triplet gap of 0.26 eV (vertical) and 0.25 eV (adiabatic). Although such BS calculations have been achieved for one isomer only, we may expect that for all the other isomers with similar electronic structure the BS ms=0 solution to be lower in energy than the singlet closed shell one, and quite close to that of the triplet spin-state.

In the case of the $AL^+$ non-planar molecular ion, a positive singlet-triplet (S-T) gap in the range [0.49 - 0.78 eV] is also found for most of the $AL^+$-H isomers. This implies that for these cases the triplet-spin states are more stable than the closed-shell ones, as expected from theoretical and experimental studies on dehydrogenated PAHs [23]. The triplet-state dissociation energies reside in the [4.55 - 4.89 eV] range. Nevertheless, we have found two geometries in which the S-T gap is negative. These correspond to the removal of a hydrogen from sterically hindered sites ($AL^+$-$H_f$ and $AL^+$-$H_g$) for which S-T gap values of -1.15 and -0.95 eV respectively have been found. Moreover, the optimization of the closed shell singlet spin state for $AL^+$-$H_f$ and -$H_g$ leads to the formation of an additional intramolecular C-C bond to form a 5-membered ring cycle. Other authors have reported a facile 5-membered ring formation in PAHs [26]. In this case, this results in a very stable isomer, leading to an H-loss dissociation channel as low as of 3.43 eV. We may notice that planarity is not achieved yet, as seen in Figure 5. Several mechanisms to form this 5-membered ring product could be invoked: it could start by H-migration followed by ring closure and H loss, or involve direct H loss and ring closure, either sequential or concerted. Unraveling such mechanisms is challenging, in particular due to the potential multireference character of the electronic states coming into play and possible spin-state crossing, and this is out of the scope of the present paper.

To summarize, considering only the thermodynamics of the reaction, the most energetically favorable paths for the loss of H in both isomers involve bay hydrogens. In the case of $AE^+$ the most stable product is a triplet-spin state ($\Delta H(0K)$= 4.67 eV). In addition, in $AL^+$, the particular local edge environment of the $H_f$ and $H_g$ sites favors the formation of a particularly stable closed-shell singlet spin state with a 5-membered ring, reducing the dissociation energy considerably by 1.24 eV (ΔH(0K)= 3.43 eV).

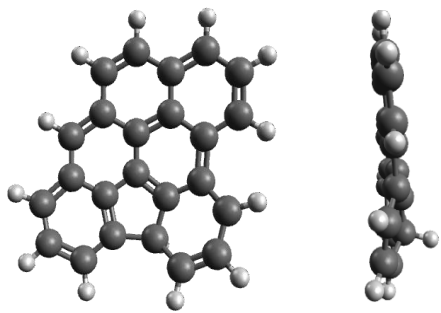

**Figure 5. Front and side view of the optimized geometry of AL$^+$ after the removal of H$_f$.**

### 3.2.2. H+H loss

In order to shed light on the double dehydrogenation mechanisms, we first assumed a sequential H loss after the loss of H$_f$ from both molecules. As discussed before, this particular position involves the lowest dissociation energies, especially for AL$^+$ where a 5-membered ring was formed in the singlet state. A second hydrogen was subsequently removed from all the remaining positions, and the theoretical dissociation energy for 2 consecutive H losses was estimated. In the following, we focus on the most energetically favorable paths for each isomer. Detailed results are reported in Table 2 of the Annex.

In the case of AE$^+$ we observe no significant difference in the values regardless of the position or spin multiplicity after the second hydrogen is removed (after H$_f$). In all cases, around 5 eV more are needed to extract this second hydrogen, adding to the 4.67 eV energy required for the first dehydrogenation in the triplet state. Previous works have estimated that the double dehydrogenation of PAHs is more favorable if the hydrogens involved are adjacent [27]. Since H$_f$ does not have any adjacent hydrogen, we also studied also the initial removal of H$_g$, which requires about the same energy (+ 0.04 eV at our level of theory, see Table 1 in the Annex), followed by a subsequent subtraction of H$_h$, or H$_i$, or H$_k$. This most favorable geometry is the doublet spin-state AE$^+$-H$_g$-H$_h$ corresponding to an addition of 3.84 eV above the precursor AE$^+$-H$_g$, or 8.55 eV above the initial reactant AE$^+$. It is important to note that this favorable geometry is found in other sites of AE$^+$, like (AE$^+$-H$_b$-H$_c$), (AE$^+$-H$_k$-H$_l$) or (AE$^+$-H$_n$-H$_a$).

Regarding AL$^+$, the lowest theoretical dissociation energy corresponds to the loss of H$_g$ from AL$^+$-H$_f$, leading to a final fully-planar structure maintaining the 5-membered ring formed by the loss of H. This reaction is barrierless. The final doubly dehydrogenated species (AL$^+$-H$_f$-H$_g$) lies at only 5.39 eV above AL$^+$ (i.e., 1.96 eV above AL$^+$-H$_f$). The remaining AL$^+$-H$_f$-H$_x$ positions (x= a ,..., e, h, ..., n) are less energetically favorable than AL$^+$-H$_f$-H$_g$, and, in general, in agreement with the values found for the double dehydrogenation of AE$^+$. In brief, DFT calculations suggest that the loss of two consecutive hydrogens would require a critical energy input of 8.55 eV in the case of AE$^+$ and 5.96 for AL$^+$. This translates in an energy difference of 2.59 eV.

### 3.2.3. H$_2$ loss

Similarly to other authors, for the naphthalene and coronene cations [21], [12], we assumed a first H migration to the adjacent position followed by a subsequent H$_2$ loss. This is illustrated in Figure 6. In the manuscript, and more precisely in Figures 7 and 8, are reported the lowest energetic paths that have been determined for the two isomers. We specify that only doublet spin-state potential energy surfaces (PES) were investigated. Regarding the first step (H migration), both clockwise and anticlockwise motions were explored (see Tables 3 and 4 in the Annex).

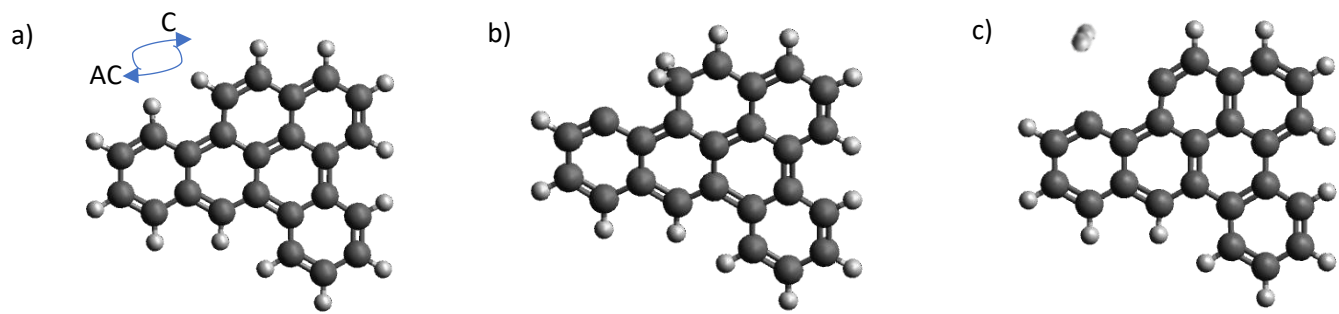

**Figure 6.** Scheme illustrating the elementary steps computed for the loss of $H_2$ (two-step mechanism involving bay H). a) The arrows represent the clockwise (C) and anticlockwise (AC) migration. b) Configuration after clockwise H-migration ($H_b$ migration to the carbon bearing $H_a$, leading to a $sp^3$ configuration). c) Final geometry with the loss of $H_2$. $AE^+$ was used as an example.

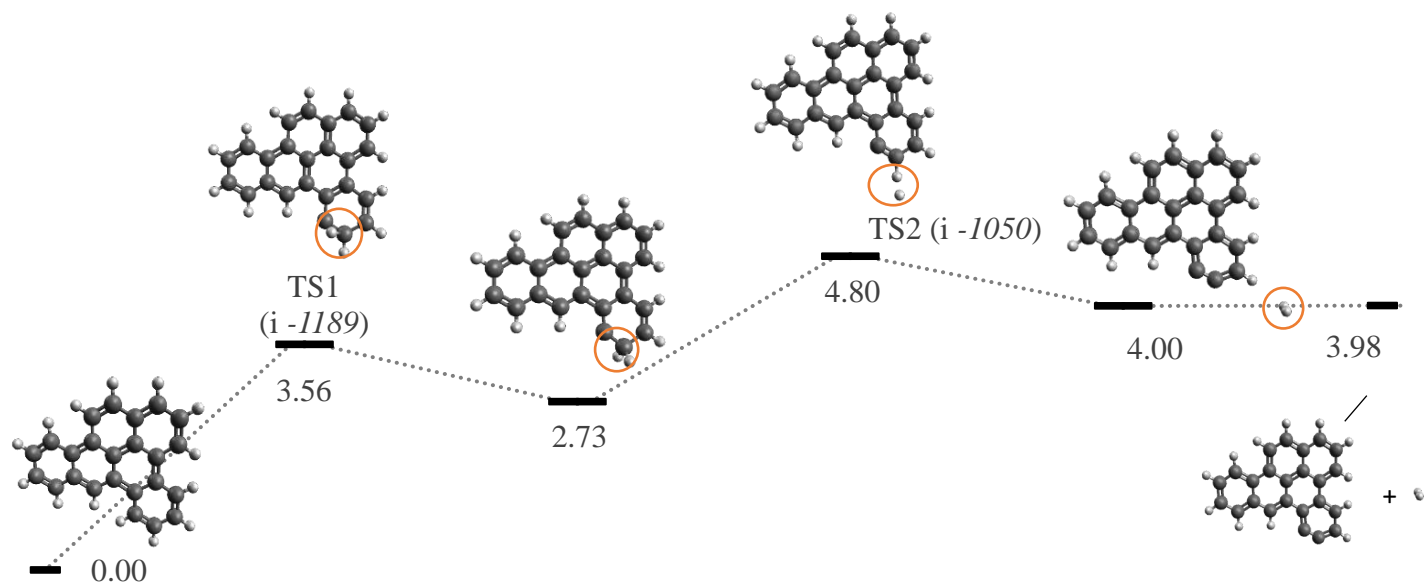

**Figure 7.** Elementary steps (B3LYP/6-31G(d,p), doublet spin-state PES) of the more favorable path leading to the formation of $H_2$ from $AE^+$, involving the lowest energy TS. The relative energy values with respect to the reactant, including ZPE, are given in eV. The imaginary frequencies of the TS are shown in italics and given in cm$^{-1}$.

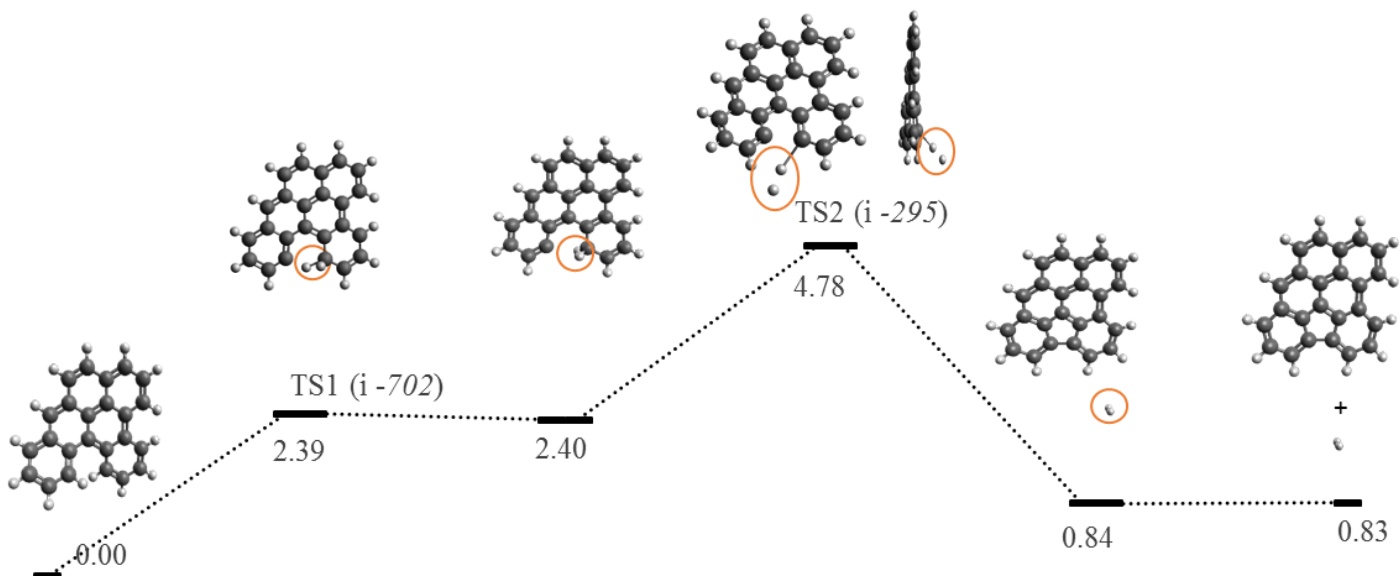

**Figure 8.** Elementary steps (B3LYP/6-31G(d,p), doublet spin-state PES) of the more favorable path leading to the formation of H$_2$ from AL$^{+\cdot}$. The lowest energy transition states involve the special hydrogens H$_f$ and H$_g$. The relative energy values with respect to the reactant, including ZPE, are given in eV. The imaginary frequencies of the TS are shown in italics and given in cm$^{-1}$.

In the case of AE$^+$ the most favorable pathway corresponds to the loss of two hydrogens in the same ring, similarly to the H+H channel (see Figure 7). The final products lie 3.98 eV above the reactant. The barrier, found at 4.80 eV, corresponds to the transition state towards the loss of the two hydrogen atoms attached to the same carbon. The values obtained for the hydrogen-migrated geometry (2.73 eV), and for the associated TS1 (3.56 eV) are in line with the ones calculated elsewhere for the coronene cation [28]. For AL$^+$ the transition state corresponding to the migration of H$_f$ onto the carbon bearing H$_g$ (TS1) effectively disappears after ZPE correction, lying at a similar energy as the optimized geometry with both hydrogens (2.4 eV). This suggests that this migration occurs without a reverse barrier (see Figure 8). The barrier of the reaction is located at 4.78 eV and it corresponds to the loss of H$_2$, similarly to AE$^+$. The final 5-membered-ring geometry lies only 0.83 eV above the parent isomer. Further, assuming a second consecutive H migration to the following carbon, we obtain a transition state of 3.26 eV and a geometry after this migration of 2.62 eV (Figure 9). This suggests that after the first migration, subsequent migrations may occur, resulting in a higher H$_2$ dissociation time than the one required for a direct H loss.

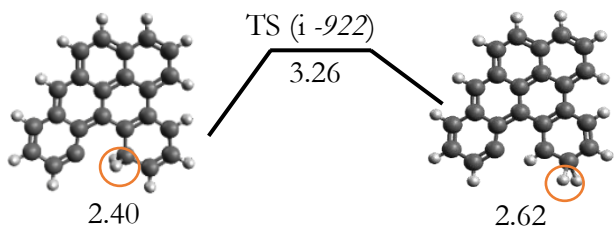

**Figure 9.** Second H migration for AL$^+$ in the path towards H$_2$ loss. The energy required for this second migration (around 3.3 eV) is lower than the barrier of the reaction (4.80 eV).

## 4. DISCUSSION

The DFT calculations provide a critical (minimal) energy for the fragmentation to occur. Experimental values are those observed at the timescale of the experiments. In our case, the threshold would be associated to the timescale of collisional energy relaxation. In addition, as described in Section 3.1, we derived onset energies above this threshold at a given observed abundance of the fragments. Assuming that the dissociation occurs from the ground electronic state, additional energy compared to the critical energy is expected to be stored in non-dissociating modes following intramolecular vibrational redistribution. The larger the molecule is, the larger the amount of stored energy will be. Since the two species of study are isomers we expect that, for a similar critical energy, the total energy stored in the molecule to achieve dissociation on experimental timescales will be similar. Following these statistical considerations, we can use the results of our DFT calculations to rationalize our experimental findings on $AE^+$ and $AL^+$.

Regarding H loss, the difference in experimental onset energies between $AE^+$ and $AL^+$ is around 2.2 eV (Figure 4a). DFT calculations assuming a direct H loss confirm the existence of two energetically favorable positions in $AL^+$: $H_f$ and $H_g$. The steric hindrance in these positions, which forces the molecule out of plane, causes $AL^+$ to lose H from these sites at considerably lower energies (1.2 eV) than $AE^+$ does from any site, forming a singlet-spin state molecular cation with a 5-membered ring. The additional 1 eV difference between the experimental and theoretical dissociation energies can be attributed to the part of the energy that is stored in the non-dissociating modes. This suggests that direct H loss is a good explanation for the single dehydrogenation mechanism.

In the case of $2H/H_2$ loss, the experimental results (Figure 4b) show that the $2H/H_2$ channel opens at higher energies than the H channel by 1.3 eV and 0.6 eV for $AL^+$ and $AE^+$, respectively. Between the two isomers the experimental energy difference is 1.5 eV for this channel. Resolving the specific mechanism that drives the double dehydrogenation at low energies has proven to be a challenge. In our DFT calculations, we assumed two possible scenarios: (i) two successive H losses (H+H) and (ii) H migration followed by a consequent $H_2$ loss (see sections 3.2.2 and 3.2.3). In the first case, computing just thermodynamical calculations we found that the loss of H+H requires at least 2.6 eV more energy in $AE^+$ than in $AL^+$. We note that, except for the case of the 5-membered ring formation in $AL^+$, the most energetically favorable path for the sequential loss of two H from the energetic point of view is the loss of two adjacent hydrogens from the same ring, in which the first H removal relieves some steric hindrance. The overall process requires at least 6 eV and 8.6 eV for $AL^+$ and $AE^+$, respectively. In the case of $H_2$ loss, we found more energetically favored channels with maximum energies of around 5 eV.

From these energetic considerations we can conclude that competition between $H_2$ and 2H channels could be possible for $AL^+$ starting at about 9.3 eV. In the case of $AE^+$, the large energy (8.6 eV) required for the 2H loss would be difficult to reconcile with an absorbed energy of ~10 eV, considering that some of it will be stored in the other non-dissociating modes. Therefore we likely observe the opening of the $H_2$ channel at ~10.8 eV. To conclude we can rationalize the difference in the experimental curves observed in Figure 4b by invoking the role of the special H of $AL^+$ ($H_f$ and $H_g$) in opening lowest energy dissociation channels.

A warning on these results remains on the accuracy in the description of the ground-state electronic structure for the dehydrogenated PAH cations. As was mentioned in section 3.2.1., open-shell electronic structures come into play and they may become difficult to describe at the DFT level when "low spin" multiple radical species compete with the "high spin" ones, the latter being overall correctly described at the DFT level. In addition, more complicated pathways leading to the loss of H and $2H/H_2$ may come into play, and this will be explored with molecular dynamics simulations, as achieved recently for compact PAHs [20].

## 5. CONCLUSIONS

Several studies have explored the effect of structure ([29–31]), size ([32,33]) and hydrogenation state ([34], [14]) in the photo-dissociation of PAHs. Nevertheless, dehydrogenation pathways for non-compact PAHs have not been thoroughly assessed. In this work we present the first study of VUV photo-fragmentation of two dibenzopyrene cation isomers that contain bay areas and differ in 3D structure (while $AE^+$ is fully planar, $AL^+$ presents a non-planar region).

Our experimental results show a small value (around 1 eV) in the energy shift between H loss and 2H/$H_2$ loss. DFT calculations reveal low energy dissociation paths involving bay hydrogens for both isomers. Moreover, all dehydrogenation channels occur at lower energies for $AL^+$ compared to $AE^+$. This was rationalized by the theoretical results, which indicate that $AL^+$ displays more energetically favorable paths for H and H+H loss due to the existence of two hydrogens in the non-planar part of the molecule. The loss of these hydrogens leads to both the relief of the steric constraint and the formation of a 5-membered ring isomer, more stable than any other dissociation products. In the case of $AE^+$ the $H_2$ channel is found to be favored because the 2H channel is open at an energy that is too close to the dissociation threshold observed in experiments. The conclusions presented in this work, and their application to a larger population of non-compact molecules, have to be further investigated. This could impact our view of the role of PAHs on the formation of $H_2$ in space.

**ACKNOWLEDGEMENTS**

We acknowledge support from the European Research Council under the European Union's Seventh Framework Programme ERC-2013-SyG, Grant Agreement n. 610256 NANOCOSMOS. The authors would like to acknowledge L. Nahon, A. Giuliani, J. Zhen, H. Sabbah, S. Martin, G. Mulas, J.P. Champeaux and P. Mayer for their collaboration within the framework of the SOLEIL proposal # 20141153. A. Simon thanks G. Trinquier for helpful discussions regarding the broken symmetry calculations. They also acknowledge the computing facility CALMIP for allocation of computer resources. Finally, the authors thank the anonymous reviewers for their valuable comments and suggestions to improve the quality of the manuscript.

# ANNEX

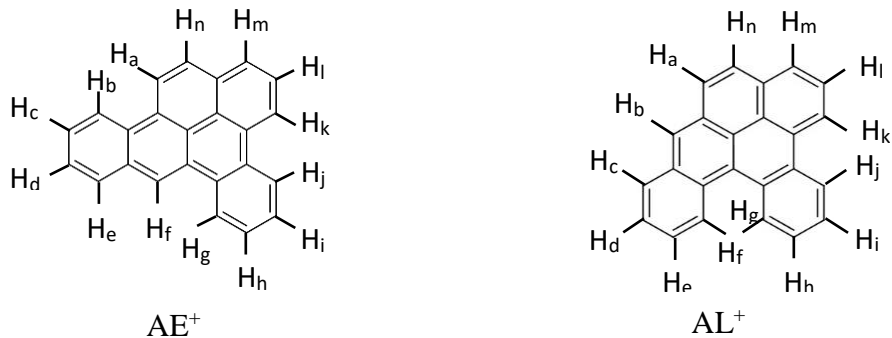

AE⁺      AL⁺

**Table 1: DFT (B3LYP/6-31 G(d,p)) calculated dissociation energies for the direct loss of hydrogen in the different positions. Calculated $\widehat{S}^2$ eigenvalues for the triplet spin-states are indicated in parenthesis. Spin contamination occurs if the values are far from 2.0 in the case of the triplet.**

|  | ΔH(0K) AE⁺ [eV] | | ΔH(0K) AL⁺ [eV] | |
| --- | --- | --- | --- | --- |
|  | Triplet | Singlet | Triplet | Singlet |
| $C_{24}H_{14}^+ - H_a$ | 4.77 (2.02) | 5.50 | 4.85 (2.02) | 5.38 |
| $C_{24}H_{14}^+ - H_b$ | 4.75 (2.02) | -    | 4.79 (2.02) | 5.43 |
| $C_{24}H_{14}^+ - H_c$ | 4.82 (2.03) | -    | 4.82 (2.03) | 5.45 |
| $C_{24}H_{14}^+ - H_d$ | 4.89 (2.02) | -    | 4.89 (2.03) | 5.48 |
| $C_{24}H_{14}^+ - H_e$ | 4.82 (2.04) | -    | 4.82 (2.03) | 5.32 |
| $C_{24}H_{14}^+ - H_f$ | 4.67 (2.03) | 5.38 | 4.58 (2.02) | 3.43 |
| $C_{24}H_{14}^+ - H_g$ | 4.71 (2.03) | -    | 4.55 (2.03) | 3.60 |
| $C_{24}H_{14}^+ - H_h$ | 4.85 (2.03) | -    | 4.86 (2.02) | 5.43 |
| $C_{24}H_{14}^+ - H_i$ | 4.84 (2.03) | -    | 4.83 (2.03) | 5.53 |
| $C_{24}H_{14}^+ - H_j$ | 4.73 (2.03) | -    | 4.74 (2.03) | 5.40 |
| $C_{24}H_{14}^+ - H_k$ | 4.70 (2.04) | -    | 4.67 (2.03) | 5.45 |
| $C_{24}H_{14}^+ - H_l$ | 4.88 (2.02) | -    | 4.85 (2.02) | 5.34 |
| $C_{24}H_{14}^+ - H_m$ | 4.83 (2.04) | -    | 4.81 (2.03) | 5.59 |
| $C_{24}H_{14}^+ - H_n$ | 4.82 (2.03) | 5.59 | 4.82 (2.02) | 5.44 |

**Table 2: DFT (B3LYP/6-31 G(d,p)) calculated dissociation energy for two consecutive H losses for the AE⁺ and AL⁺ isomers. The spin states of the products are specified. Calculated $\widehat{S}^2$ eigenvalues are indicated in parenthesis. Spin contamination occurs if the values are far from 0.75 for doublet and 3.75 for quartet. Left table: The second dehydrogenation occurs after a first loss of H from $H_f$. Right table: Two H losses departing from the same ring in AE⁺, where the first dehydrogenation is the loss of $H_g$.**

|  | ΔH(0K) AE⁺ [eV] | | ΔH(0K) AL⁺ [eV] |
| --- | --- | --- | --- |
|  | Doublet | Quartet | Doublet |
| $C_{24}H_{14}^+ - H_f - H_a$ | -           | 9.44 (3.78) | 8.37 (0.76) |
| $C_{24}H_{14}^+ - H_f - H_b$ | 9.19 (1.68) | 9.49 (3.78) | 8.36 (0.75) |
| $C_{24}H_{14}^+ - H_f - H_c$ | 9.54 (1.78) | 9.38 (1.78) | 8.32 (0.77) |
| $C_{24}H_{14}^+ - H_f - H_d$ | 9.67 (1.78) | 9.50 (1.78) | 8.34 (0.76) |
| $C_{24}H_{14}^+ - H_f - H_e$ | 9.70 (1.77) | 9.53 (1.77) | 8.34 (0.77) |
| $C_{24}H_{14}^+ - H_f - H_g$ | 9.64 (1.78) | 9.42 (3.79) | 5.96 (0.77) |
| $C_{24}H_{14}^+ - H_f - H_h$ | 9.66 (1.78) | 9.49 (1.77) | 8.34 (0.77) |
| $C_{24}H_{14}^+ - H_f - H_i$ | 9.58 (1.65) | 9.64 (3.79) | 8.32 (0.77) |
| $C_{24}H_{14}^+$ - $H_f$ - $H_j$ | 9.53 (1.78) | 9.70 (3.79) | 8.28 (0.80) |
| $C_{24}H_{14}^+$ - $H_f$ - $H_k$ | 9.45 (1.78) | 9.71 (3.80) | 8.31 (0.76) |
| $C_{24}H_{14}^+$ - $H_f$ - $H_l$ | 9.64 (1.78) | 9.50 (1.78) | 8.34 (0.76) |
| $C_{24}H_{14}^+$ - $H_f$ - $H_m$ | 9.71 (1.76) | 9.54 (3.80) | 8.38 (0.76) |
| $C_{24}H_{14}^+$ - $H_f$ - $H_n$ | 9.68 (1.77) | 9.49 (3.79) | 8.34 (0.76) |

| | $\Delta H(0K)$ AE$^+$ [eV] | |
|---|---|---|
| | Doublet | Quartet |
| $C_{24}H_{14}^+$ - $H_g$ - $H_h$ | 8.55 (0.77) | 9.92 (3.77) |
| $C_{24}H_{14}^+$ - $H_g$ - $H_i$ | 8.75 (0.77) | 9.67 (3.79) |
| $C_{24}H_{14}^+$ - $H_g$ - $H_j$ | 9.52 (1.78) | 9.52 (3.77) |

**Table 3:** B3LYP/6-31G(d,p) doublet spin state minima and transition states (TS) computed on the potential energy surface related to the H$_2$ loss from AE$^+$. AC stands for anticlockwise and C for clockwise movement. Values (in eV) are given with respect to the reactant. Raw energy values are followed by ZPE in parenthesis. The dashes represent failed (unconverged) calculations.

| AE$^+$ | TS1 | H-shifted | TS2 | M$^+$ ---H$_2$ | [M$^+$-H$_2$] + H$_2$ |
|---|---|---|---|---|---|
| H$_a$- H$_b$ (AC) | - | 2.79 (-0.07) | - | 5.52 (-0.44) | 5.53 (-0.44) |
| H$_a$- H$_b$ (C) | - | 2.84 (-0.07) | - | 5.45 (-0.41) | 5.53 (-0.44) |
| H$_b$- H$_c$ (AC) | 3.54 (-0.17) | 2.47 (-0.05) | 5.69 (-0.33) | 4.43 (-0.39) | 4.43 (-0.41) |
| H$_b$- H$_c$ (C) | 3.68 (-0.17) | 2.85 (-0.07) | 5.69 (-0.33) | 4.43 (-0.39) | 4.43 (-0.41) |
| H$_c$- H$_d$ (AC) | 3.94 (-0.17) | 2.98 (-0.07) | 5.65 (-0.32) | 4.66 (-0.39) | 4.67 (-0.42) |
| H$_c$- H$_d$ (C) | - | 2.64 (-0.06) | - | 4.66 (-0.39) | 4.67 (-0.42) |
| H$_d$- H$_e$ (AC) | 4.01 (-0.15) | 2.61 (-0.06) | 5.65 (-0.33) | 4.60 (-0.39) | 4.61 (-0.42) |
| H$_d$- H$_e$ (C) | 3.68 (-0.17) | 2.92 (-0.07) | 5.44 (-0.30) | 4.60 (-0.39) | 4.61 (-0.42) |
| H$_e$- H$_f$ (AC) | 4.35 (-0.20) | 2.16 (-0.03) | 5.57 (-0.30) | 6.04 (-0.44) | 5.57 (-0.44) |
| H$_e$- H$_f$ (C) | 4.65 (-0.16) | 2.50 (-0.05) | 5.44 (-0.42) | 5.56 (-0.42) | 5.57 (-0.44) |
| H$_f$- H$_g$ (AC) | 3.62 (-0.19) | 2.81 (-0.07) | 5.74 (-0.41) | 5.50 (-0.41) | 5.51 (-0.43) |
| H$_f$- H$_g$ (C) | 3.08 (-0.17) | 2.06 (-0.04) | 5.30 (-0.42) | 5.30 (-0.40) | 5.51 (-0.43) |
| H$_g$- H$_h$ (AC) | 3.75 (-0.19) | 2.80 (-0.07) | 5.22 (-0.42) | 4.39 (-0.39) | 4.40 (-0.41) |
| H$_g$- H$_h$ (C) | 3.86 (-0.18) | 2.94 (-0.07) | 5.76 (-0.33) | 4.38 (-0.38) | 4.40 (-0.41) |
| H$_h$- H$_i$ (AC) | 3.93 (-0.18) | 2.92 (-0.07) | - | 4.67 (-0.39) | 4.68 (-0.42) |
| H$_h$- H$_i$ (C) | 3.93 (-0.18) | 2.95 (-0.07) | - | 4.67 (-0.39) | 4.68 (-0.41) |
| H$_i$- H$_j$ (AC) | 3.85 (-0.18) | 2.95 (-0.07) | - | 4.42 (-0.39) | 4.41 (-0.41) |
| H$_i$- H$_j$ (C) | 3.73 (-0.18) | 2.78 (-0.06) | 5.25 (-0.42) | 4.42 (-0.39) | 4.42 (-0.41) |
| H$_j$- H$_k$ (AC) | - | 2.29 (-0.05) | - | 5.36 (-0.41) | 5.52 (-0.44) |
| H$_j$- H$_k$ (C) | - | 2.85 (-0.08) | - | 5.51 (-0.72) | 5.52 (-0.44) |

Table 4: B3LYP/6-31G(d,p) doublet spin state minima and transition states (TS) computed on the potential energy surface related to the H$_2$ loss from AE$^+$. AC stands for anticlockwise and C for clockwise movement. Values (in eV) are given with respect to the reactant. Raw energy values are followed by ZPE in parenthesis. The dashes represent failed (unconverged) calculations.

| AL$^+$ | TS1 | H-shifted | TS2 | M$^+$ ---H$_2$ | [M$^+$-H$_2$] + H$_2$ |
|---|---|---|---|---|---|
| H$_a$- H$_b$ (AC) | - | 2.12 (-0.03) | 5.72 (-0.37) | 5.57 (-0.41) | 5.57 (-0.43) |
| H$_a$- H$_b$ (C) | - | 2.87 (-0.05) | - | 6.16 (-0.41) | 5.57 (-0.43) |
| H$_b$- H$_c$ (AC) | 4.52 (0.10) | 2.59 (-0.05) | - | 5.70 (-0.41) | 5.71 (-0.43) |
| H$_b$- H$_c$ (C) | - | 2.09 (-0.03) | - | 5.56 (-0.41) | 5.64 (-0.43) |
| H$_c$- H$_d$ (AC) | 3.67 (-0.17) | 2.91 (-0.07) | 5.66 (-0.33) | 4.59 (-0.39) | 4.60 (-0.42) |
| H$_c$- H$_d$ (C) | 3.66 (-0.17) | 2.55 (-0.05) | 5.42 (-0.43) | 4.59 (-0.38) | 4.60 (-0.42) |
| H$_d$- H$_e$ (AC) | 3.97 (-0.18) | 2.63 (-0.06) | - | 4.66 (-0.40) | 4.66 (-0.42) |
| H$_d$- H$_e$ (C) | 3.99 (-0.18) | 2.98 (-0.07) | - | 4.66 (-0.39) | 4.66 (-0.42) |
| H$_e$- H$_f$ (AC) | 3.71 (-0.17) | 2.90 (-0.06) | - | 4.22 (-0.41) | 4.23 (-0.42) |
| H$_e$- H$_f$ (C) | 3.38 (-0.19) | 2.35 (-0.07) | 5.30 (-0.35) | 4.22 (-0.41) | 4.23 (-0.42) |
| H$_f$- H$_g$ (AC) | 2.57 (-0.18) | 2.48 (-0.08) | 5.22 (-0.44) | 1.15 (-0.31) | 1.16 (-0.33) |
| H$_f$- H$_g$ (C) | - | 2.49 (-0.08) | - | 1.15 (-0.31) | 1.16 (-0.33) |
| H$_g$- H$_h$ (AC) | 3.53 (-0.19) | 2.72 (-0.10) | 5.35 (-0.35) | 4.21 (-0.39) | 4.22 (-0.42) |
| H$_g$- H$_h$ (C) | - | 2.88 (-0.06) | - | 4.21 (-0.40) | 4.22 (-0.42) |
| H$_h$- H$_i$ (AC) | - | 2.99 (-0.07) | - | 4.65 (-0.39) | 4.66 (-0.42) |
| H$_h$- H$_i$ (C) | - | 2.83 (-0.07) | - | 4.65 (-0.45) | 4.66 (-0.42) |
| H$_i$- H$_j$ (AC) | 3.82 (-0.18) | 2.97 (-0.07) | - | 4.40 (-0.39) | 4.41 (-0.41) |
| H$_i$- H$_j$ (C) | 3.74 (-0.18) | 2.69 (-0.06) | 5.32 (-0.30) | 4.40 (-0.38) | 4.41 (-0.41) |
| H$_j$- H$_k$ (AC) | - | 2.31 (-0.05) | - | 4.71 (-0.36) | 4.72 (-0.38) |
| H$_j$- H$_k$ (C) | - | 2.88 (-0.07) | - | 5.50 (-0.42) | 5.51 (-0.44) |